\documentclass[lettersize,journal]{IEEEtran}
\usepackage{amsmath,amsfonts}
\usepackage{algorithmic}
\usepackage{algorithm}
\usepackage{array}
\usepackage[caption=false,font=normalsize]{subfig}
\usepackage{textcomp}
\usepackage{stfloats}
\usepackage{url}
\usepackage{verbatim}
\usepackage{graphicx}
\usepackage{cite}
\usepackage{xcolor}
\usepackage{subfig}
\usepackage[labelsep=colon
]{caption}
\usepackage{flushend} 

\newcommand{\mycomment}[1]{}

\begin{document}

\title{Dual mode multispectral imaging system for food and agricultural product quality estimation}

\author{Darsha Udayanga, Ashan Serasinghe, Supun Dassanayake, Roshan Godaliyadda,~\IEEEmembership{Senior Member,~IEEE}, \\Vijitha Herath,~\IEEEmembership{Senior Member,~IEEE}, Mervyn Parakrama Ekanayake,~\IEEEmembership{Senior Member,~IEEE}, Pasindu Malshan
\thanks{The authors are with the Department of Electrical and Electronic Engineering, Faculty of Engineering, University of Peradeniya, Sri Lanka.}
\thanks{Corresponding author's email: darsha@eng.pdn.ac.lk}}



\maketitle

\begin{abstract}
Multispectral imaging coupled with Artificial Intelligence, Machine Learning and Signal Processing techniques work as a feasible alternative for laboratory testing, especially in food quality control. Most of the recent related research has been focused on reflectance multispectral imaging but a system with both reflectance, transmittance capabilities would be ideal for a wide array of specimen types including solid and liquid samples.  In this paper, a device which includes a dedicated reflectance mode and a dedicated transmittance mode is proposed. Dual mode operation where fast switching between two modes is facilitated. An innovative merged mode is introduced in which both reflectance and transmittance information of a specimen are combined to form a higher dimensional dataset with more features. Spatial and temporal variations of measurements are analyzed to ensure the quality of measurements.  The concept is validated using a standard color palette and specific case studies are done for standard food samples such as turmeric powder and coconut oil proving the validity of proposed contributions. The classification accuracy of standard color palette testing was over 90\% and the accuracy of coconut oil adulteration was over 95\%. while the merged mode was able to provide the best accuracy of 99\% for the turmeric adulteration. A linear functional mapping was done for coconut oil adulteration with an R\textsuperscript{2} value of 0.9558.

\end{abstract}


\section{Introduction}
\IEEEPARstart{F}{ood} quality is an integral and essential part of global food security. Ensuring food quality includes the assurance that the food is void of harmful contaminants and unacceptable adulterants \cite{gen_2,gen_1}. However, analyzing food samples for contaminants and adulterants requires sophisticated laboratory measurement systems and relies on skilled professionals. There are hardly any convenient and robust measurement systems that can be used at field level even for the most rudimentary of such tests.  

Humans rely on their senses, especially smell and vision to gauge the potential quality of food prior to consumption. As the sense of smell involves complex chemical phenomena, it does not provide an easy avenue for automatic detection. On the other hand, visual sensation provides ample opportunities for such automated testing. Therefore, vision based sensing is a popular choice for quality assessment in various applications \cite{Cite_1, Cite_2, Cite_3}, including but not limited to medical \cite{Cite_7,jist_4}, astronomical \cite{Cite_8,ros_1}, cultural heritage \cite{i2m_2,i2m_4} and agricultural \cite{Cite_9,jist_1} fields.

Multispectral Imaging (MSI) is an enhancement of tri-color red-green-blue (RGB) imaging. MSI utilizes multiple narrow bands of color covering not only the visual range but also near-infrared (NIR) and near-ultraviolet (NUV). Therefore, MSI provides a rich set of information \cite{Cite_4}, over standard RGB imaging, including finer details that depend on chemical properties. This advantage of MSI can be further enhanced via the incorporation of the recent advances in artificial intelligence (AI) and machine learning (ML) \cite{gen_3,els_2}.

There are high-end industrial grade MSI devices, which are usually purpose-specific, bulky, and expensive \cite{jist_2,els_5,ros_2}. However, in cases where onsite testing is necessary, opting for a versatile and portable design with innovative concepts to enhance and enrich the captured MSI is a more reasonable approach. Furthermore, since the purpose of such a system is primarily to be deployed in the large scale and multilayered retail market, simplicity and cost effectiveness take priority. Moreover, the use of MSI system may be extended beyond food applications to various other types of specimens that require simple, cost-effective onsite testing. 

Contemporary MSI devices utilized different techniques to acquire multispectral images of specimens \cite{jist_3}. One study \cite{Cite_10} has utilized liquid crystal tunable filters (LCTFs) to divide the spectrum. Another design \cite{Cite_11} has used a multispectral filter array (MSFA)  and a multispectral demosaicing algorithm. Using filters is a common practice for acquiring MSI as evidenced by this device \cite{Cite_12} which comprises several mirrors and filters. An iris capture device \cite{Cite_13} was also designed using filters and LEDs as the illumination source. Two studies \cite{Cite_14, Cite_15} were found to make use of hyperspectral tunable filters and diffraction grating respectively.

One study \cite{Cite_16} published in 2020 with the title of ‘Characterization of a multispectral imaging system based on narrow bandwidth power LEDs’ which has utilized an array of Narrowband LEDs with the wavelength ranging from 410 nm to 950 nm across 15 pairs of LEDs. The system includes a multispectral lighting system, an optical sensor, a light controller and an image capture environment. The multispectral lighting system was a circularly arranged set of LEDs controlled using the light controller. The light generated by the LEDs was shined upon the specimen and the reflectance image was captured using the camera. This process was carried out within a special image capture environment that was created for this purpose. The intensity of the light could be controlled using a PWM (Pulse Width Modulation) signal.

In principle, MSIs of a sample can be acquired either by illuminating it with light sources of specific narrowband wavelengths \cite{Cite_6} or by illuminating the sample with a broadband light source and filtering the required light frequencies using an array of narrowband filters \cite{Cite_5,i2m_3,els_1,els_3}. The former is simpler in design, robust, expandable, and overall, more economical; while the latter is more compact, yet much more expensive and less flexible. 

One limitation identified in the above mentioned system  \cite{Cite_16} was its inability to process liquid or similar samples due to the lack of a transmittance mode imaging method. This is a huge drawback because this limits the use of the said device strictly to solid specimens and reflectance mode. Few other above-mentioned studies also contained the same limitation  \cite{Cite_10 , Cite_11, Cite_12, Cite_13, Cite_14 , Cite_15}. The lack of a light directing mechanism of the said system \cite{Cite_16} is another drawback because the light generated by LEDs usually disperses around the generation source. Therefore, a light directing mechanism is required to guide the light onto the specimen. The other mentioned studies exhibit a few additional design limitations such as low spectral resolution \cite{Cite_13} (Only three color bands) and the lack of light directing mechanism onto the specimen \cite{Cite_14}. 

There were a few studies done combining different imaging modes, such as reflectance-transmittance mode \cite{review_pickles,review_insect} and reflectance-fluorescence mode \cite{review_fluorescence}. In the mentioned reflectance-transmittance mode studies, they utilized a conveyor belt-based imaging system to identify contaminated or defective fruits. However, the system was comparatively less mobile and power-consuming due to the utilized lighting system. Also, it was cooled to $-12$ $^{\circ}$C, expensively using thermoelectric cooling which is not preferable for a portable device. One drawback identified in their approach was that the spectrum is divided for reflectance and transmittance imaging. The former part of the spectrum is only imaged in reflectance mode while the latter part is only imaged in transmittance mode. Therefore, it fails to make use of the whole spectrum via both modes. A review paper has made a comprehensive review of different existing approaches in spectral imaging \cite{review_review_paper}. As mentioned in this paper, there is a lack of study for efficient, reliable, and cost-effective spectral imaging techniques. Furthermore, therein, the relevance of combined imaging modes to the industry is stressed while the need for validating these systems using real-world samples is emphasized. All of these concerns are addressed in our work.

Considering the nature of the problem at hand, an MSI system that illuminates the sample with narrowband LEDs is proposed in this paper as it is the more economical alternative to get more spectral bands in the desired range. Innovative design concepts of the proposed system include a dual mode capability for reflectance and transmittance imaging, a versatile modular architecture, an intensity control mechanism for flexibility, integrating hemispheres to ensure uniform light distribution and a controller system coupled with an easy-to-use user interface. Thereafter, context specific AI, data analytic and signal processing algorithms are developed to obtain functional relationships between contaminants/adulterants and statistical characteristics of multispectral measurement data. Furthermore, ML algorithms are used for classifications based on contamination/adulteration levels.

To evaluate the imaging system, system validation must be done. The validation process can be carried out in two methods. Either by testing standard colors \cite{ros_3} or by conducting experimental tests on real samples \cite{ros_4} to determine the applicability of the device. The study under consideration \cite{Cite_16} has only carried out the standard color test. While this gives substantial information on the functionality of the system, it’s challenging to determine the real-world applicability of the system. Hence, conducting experiments on real samples is required. Furthermore, the study \cite{Cite_16} has attempted to visualize the separation of 24 distinct colors using the imaging device. Correlation analysis of the spectral signatures of these colors was done by applying Principal Component Analysis (PCA) \cite{review_2_pca}. It was compared with a high grade Hyperspectral Imaging (HSI) system. Moreover, one of the systems under consideration \cite{Cite_10} has utilized both testing using standard lamps and experimental testing using traditional Chinese medicines as specimens. Another study \cite{Cite_11} was tested experimentally by capturing their own set of images and comparing them with standard multispectral datasets. Since the algorithm development part was one of the main objectives, two algorithms were applied to the dataset and the results were compared. In another study, \cite{Cite_12} their own synthesized tests were performed and PCA was used as well. A multispectral Iris Capture Device \cite{Cite_13} was validated by creating a dataset by imaging different subjects. They have used their own image-level fusion algorithm. 1-D log-Gabor wavelet recognition method \cite{Cite_17} was used to identify the iris images.  A multispectral skin imaging system \cite{Cite_14} was validated using real samples. However, they have only used two test samples which may have affected their validation accuracy. One other considered study \cite{Cite_14} has also failed to validate their device using actual real experimental testing. When examining \mycomment{the validation procedures and algorithms of}previous studies, very few studies have focused on building functional relationships between the parameters of the sample and the MSI parameters \cite{i2m_1}. Functional maps or regression models offer the ability to operate in a continuum as opposed to mere classification which operates only on a few designated
classes.

Major contributions of our work considering the current state of the art are as follows,

\begin{itemize}
\item{Dual mode transmittance-reflectance merged operation capable multispectral imaging setup. In which, signal processing-based dimension reduction algorithms such as PCA and Linear Discriminant Analysis (LDA) are used to optimally combine spatially and spectrally corrected transmittance and reflectance spectral bands for new feature generation for given classification problems.}
\item{In addition to the classification of samples based on their adulteration levels, formulation of adulteration estimation problems as a functional mapping or regression problem. In which, the Kullback-Leibler (KL) divergence from pure sample to adulterated sample is utilized to quantify the relationship.}
\end{itemize}


Essentially, the merged mode makes use of spectral components from both reflectance and transmittance modes, creating a dataset with a higher dimensionality. It provides additional information for dimension reduction algorithms to generate a more enriched feature vector set for AI and ML algorithms to operate on, which results in increased accuracy. As an example, turmeric powder can be captured in reflectance mode as a powder. Then, the same sample can be dissolved in water and captured in transmittance mode as a liquid, and both sets of images can be combined to enable the merged mode operation. Furthermore, some translucent liquids can also be tested via this merged mode. 

A study was carried out to assess the temporal and spatial variations of measured data to identify the optimum image acquisition area. To validate the device and proposed algorithms on standard and real-world data, first, a validation study was conducted on a standard color palette. Furthermore, a case study was conducted to classify among adulteration levels of rice flour on turmeric powder utilizing the merged mode. Finally, another case study was performed to build a functional relationship between the adulteration level of palm oil on coconut oil using the KL divergence as the metric.

\section{Multispectral Imaging system}

The multispectral imaging system encompasses several noteworthy features. The main elements are the lighting panels, the controller board, the image capture environment and the controller software including the graphical user interface (GUI). The user is connected to the system via a computer. 

\begin{figure}[b]
\centering
\subfloat[\centering \textrm{Front view}]{{\includegraphics[width=0.4\linewidth]{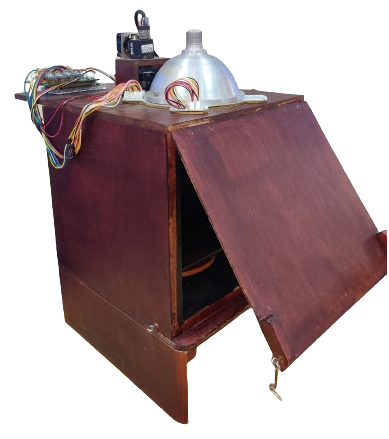} }
\label{fig_device_front}}%
\qquad
\subfloat[\centering \textrm{Rear view}]{{\includegraphics[width=0.4\linewidth]{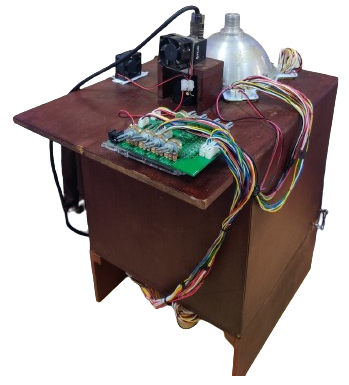} }
\label{fig_device_rear}}%
\caption{The multispectral imaging setup.}
\label{fig_device}
\end{figure}

Since the LED acts as a single point light source, the light is concentrated around the center.   Reflecting hemispheres with dull surfaces are used in combination with lighting panels to diffuse and direct the generated light in the required direction. Also, the light panels are constructed in a modular architecture, so the intensity of each light panel quarter could be handled individually, thus providing better control over the light spread and intensity on a specimen. The system covers wavelengths from Ultraviolet (UV) to NIR where the range extends from 365 nm to 940 nm.  It is made sure to have enough illumination on a specimen by using eight LEDs per color band per light panel. LEDs are distributed in a radial manner which ensures the uniformity of the light on the specimen. Finally, to establish ease of operation for the user and to maintain control over the system, the controller software is developed including a GUI which provides the user with a preview of the captured image and a lot of options in handling the system. The developed multispectral imaging setup is displayed in the Fig. \ref{fig_device}.

\begin{figure}[tb]
\centering
\includegraphics[width=0.7\linewidth]{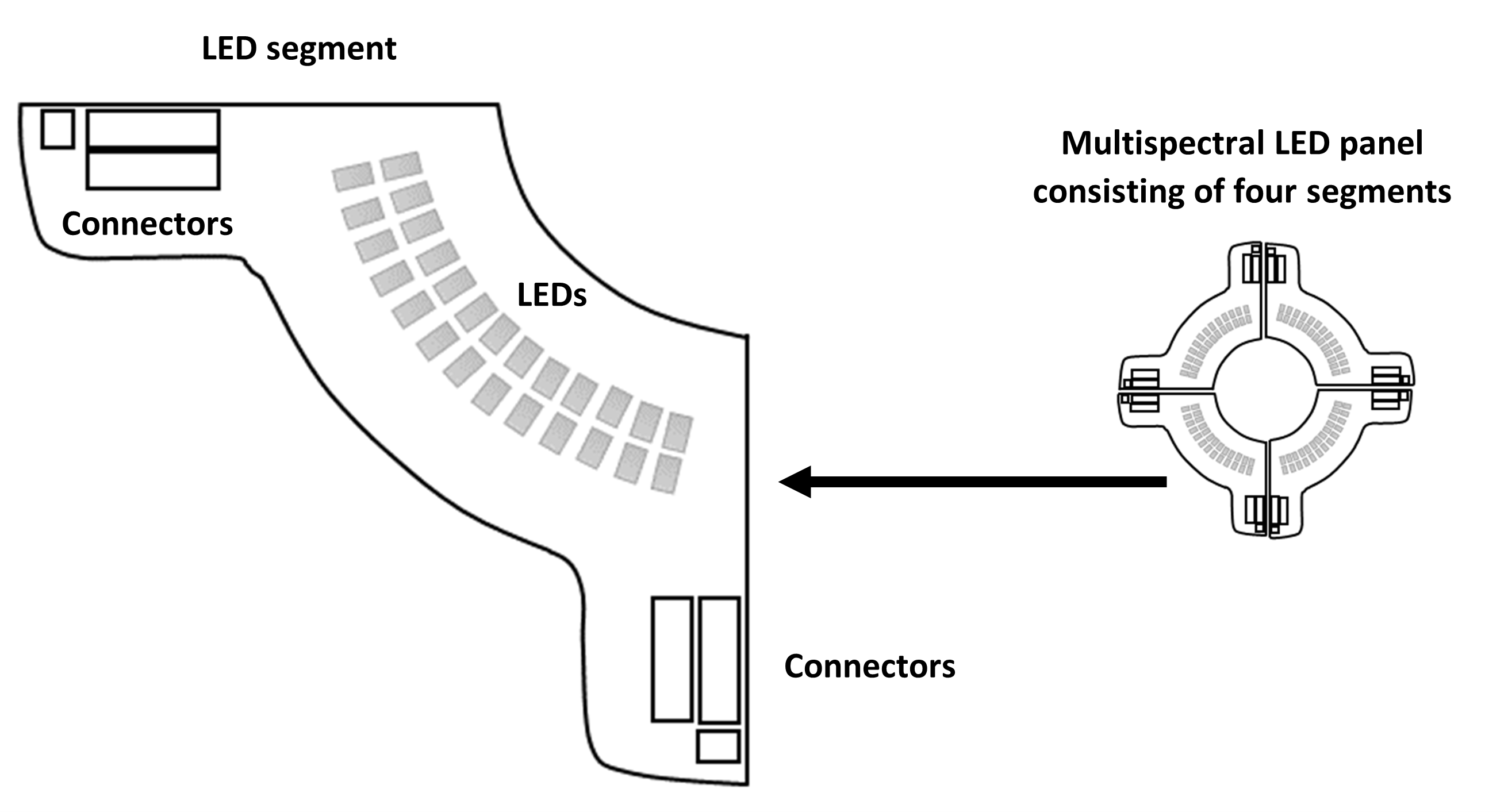}
\caption{Multispectral illumination light panel of the imaging system with one segment enlarged.}
\label{fig_light_panel}
\end{figure}

\begin{table}[tb]
    \centering
    \caption{Properties of Illumination LEDs}
    \begin{tabular}{|c|l|r|}
    \hline
        \shortstack{Dominant \\ wavelength(nm)} & \shortstack{Radiant Intensity(mW/sr) or \\Luminous Intensity(mcd)}  & \shortstack{Viewing \\Angle} \\ \hline
        365 & Not Available & 120°  \\ \hline
        405 & 2.5 mW/sr & 120°   \\ \hline
        428 & 100 mcd & 60° \\ \hline
        473 & 600 mcd & 120°   \\ \hline
        530 & 1800 mcd & 120°   \\ \hline
        575 & 70 mcd & 120°   \\ \hline
        621 & 100 mcd & 120°  \\ \hline
        660 & 150 mcd & 120°   \\ \hline
        735 & Not Available & 120°   \\ \hline
        770 & 6 mW/sr & 30°   \\ \hline
        830 & 180 mW/sr & 20°  \\ \hline
        850 & 17 mW/sr & 120°   \\ \hline
        890 & 13 mW/sr & 120°  \\ \hline
        940 & 5 mW/sr & 120°  \\ \hline
    \end{tabular}
    \label{led_table}
\end{table}

\subsection{Multispectral Light Panels}
\begin{figure}[tb]
\centering
\includegraphics[width=\linewidth]{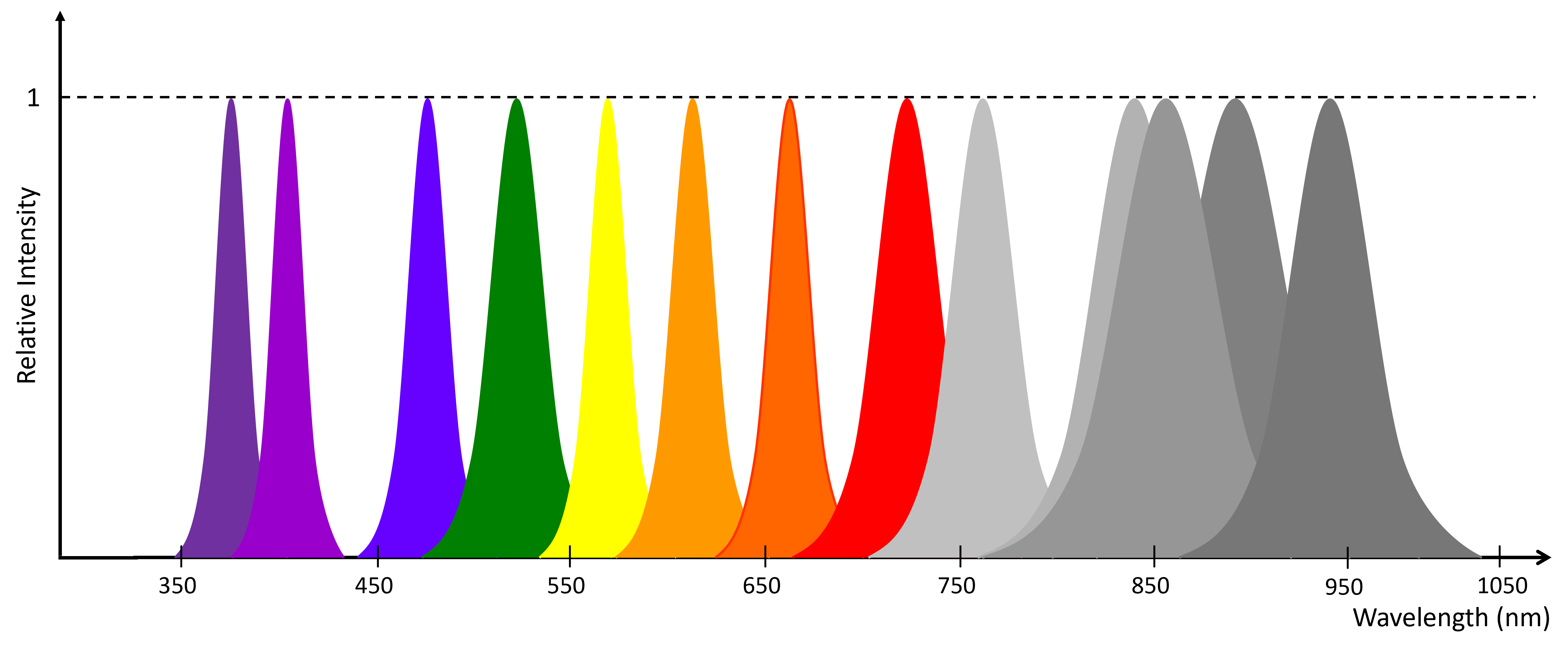}
\caption{Normalized intensity vs wavelength of LEDs.}
\label{fig_bell_curves}
\end{figure}

\begin{figure}[tb]
\centering
\includegraphics[width=0.7\linewidth]{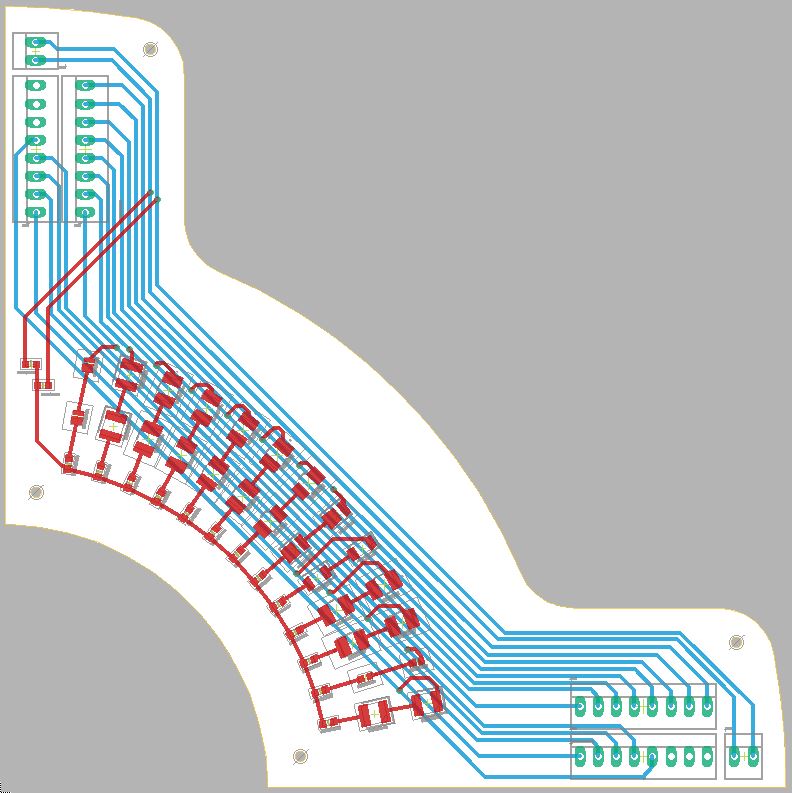}
\caption{PCB of one light segment in the LED panel.}
\label{fig_pcb_1}
\end{figure}

The LED panels are designed to have four identical segments as depicted in Fig. \ref{fig_light_panel}. Each segment has LEDs of 13 color bands with the spectral range spanning from 365 nm to 940 nm. The optical properties of the used LEDs are given in TABLE \ref{led_table}, whereas the normalized intensities of LEDs are shown in Fig. \ref{fig_bell_curves}. The device is designed in such a modular architecture to enable intensity control of each segment. This way, it permits the user to control each quarter of the light panel individually. The LEDs in each segment are placed symmetrically and radially in order to obtain a uniformly dispersed light onto the specimen. The segments are connected to each other using connectors. In addition, in case of a malfunction in one of the four segments, other segments continue to function independently. The whole system does not have to be taken down for maintenance. Other segments can function normally while the malfunctioned segment is being repaired. Malfunction identification of these light panels is also made significantly easier by deploying this modular architecture. The printed circuit board (PCB) layout of a light segment is shown in Fig. \ref{fig_pcb_1}. Although there is significant spectral overlap among certain LEDs as indicated in Fig. \ref{fig_bell_curves}, they are chosen to have different peak wavelengths from each other. Despite the high correlation between the information of these overlapping LEDs, there is still valuable uncorrelated information generated due to the differences in their peak wavelengths. Hence, the uncorrelated information is extracted by using dimensionality reduction techniques (PCA and LDA) in the feature extraction phase which will be further explained in subsequent sections \cite{review_2_lda,review_2_pca}.

\subsection{Image Capture Environment (Specimen Illumination Chamber)}

\begin{figure}[tb]
\centering
\includegraphics[width=\linewidth]{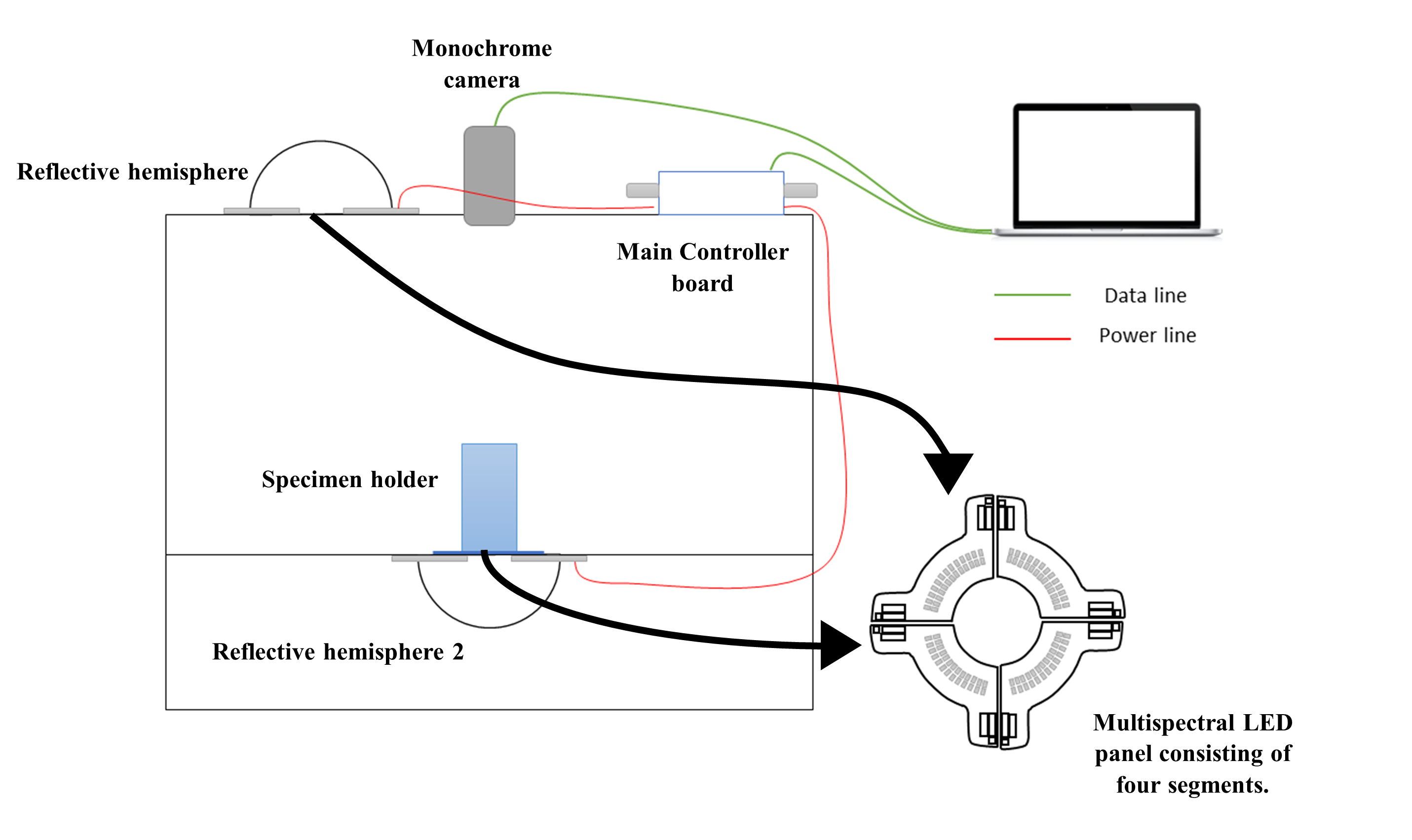}
\caption{Cross section of the imaging system.}
\label{fig_chamber}
\end{figure}

A dark chamber is used as the image capture environment for the specimens. It is an optically sealed box, and therefore, protected from any outside interference when imaging. Furthermore, the inside of the image capture environment is painted in matte black to prevent any interactions with the illuminating light. One light panel is mounted on top of the chamber for reflectance imaging purposes while the other light panel is mounted at the bottom of the chamber for transmittance imaging purposes. Each light panel is covered by a  reflecting hemisphere to ensure light dispersion and to direct the light in the intended path. A cross section of the whole system is presented in Fig. \ref{fig_chamber}

The camera is directly placed above the transmittance light panel. Therefore it can capture the light directly coming through the specimen.  Similarly, the light reflected from the specimen in the reflectance mode is also captured by the same camera. The platform which holds the specimen can be switched between a transparent glass sheet or an opaque wooden panel depending on the mode of illumination used ( transmittance or reflectance). The main controller board is mounted on the top outer surface of the chamber, closer to the camera. 

Specimen holding container should be selected to have no chemical reactions with the specimen. The specimens are usually imaged in Petri dishes made out of transparent borosilicate glass which has a low and constant light absorbance across the considered spectral range (365 nm - 940 nm) \cite{review_1_petri}.

\subsection{Control Board}
The control board is used for three main purposes. 

\begin{enumerate}
\item{Turning on/off the corresponding color band in the light panel.}
\item{Controlling the intensity of each segment of the light panel.}
\item{Communicating with the controller software.}
\end{enumerate}

\begin{figure}[b]
\centering
\includegraphics[width=0.5\linewidth]{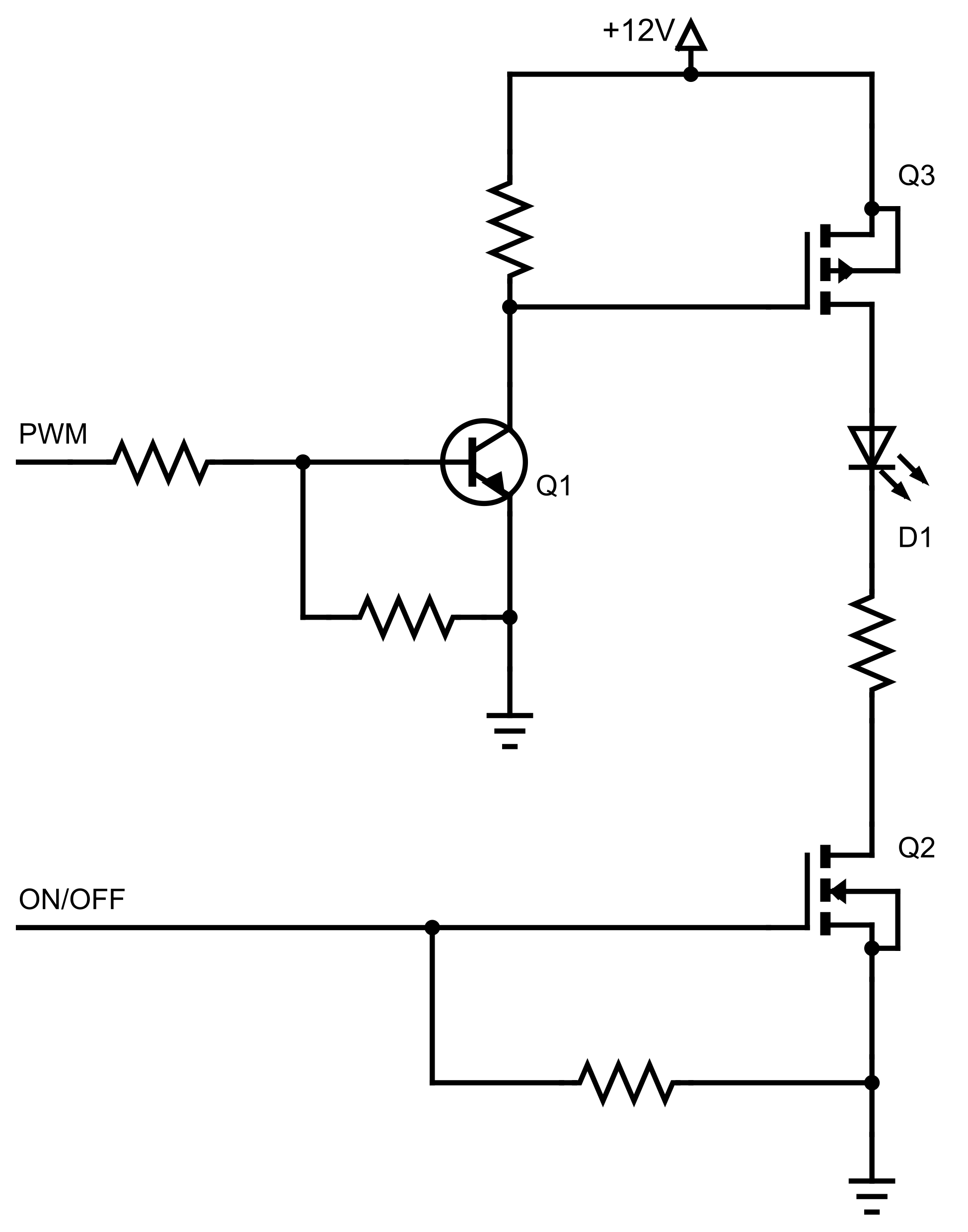}
\caption{Circuit diagram of a single color band light source.}
\label{fig_circuit}
\end{figure}

\begin{figure}[tb]
\centering
\includegraphics[width=0.7\linewidth]{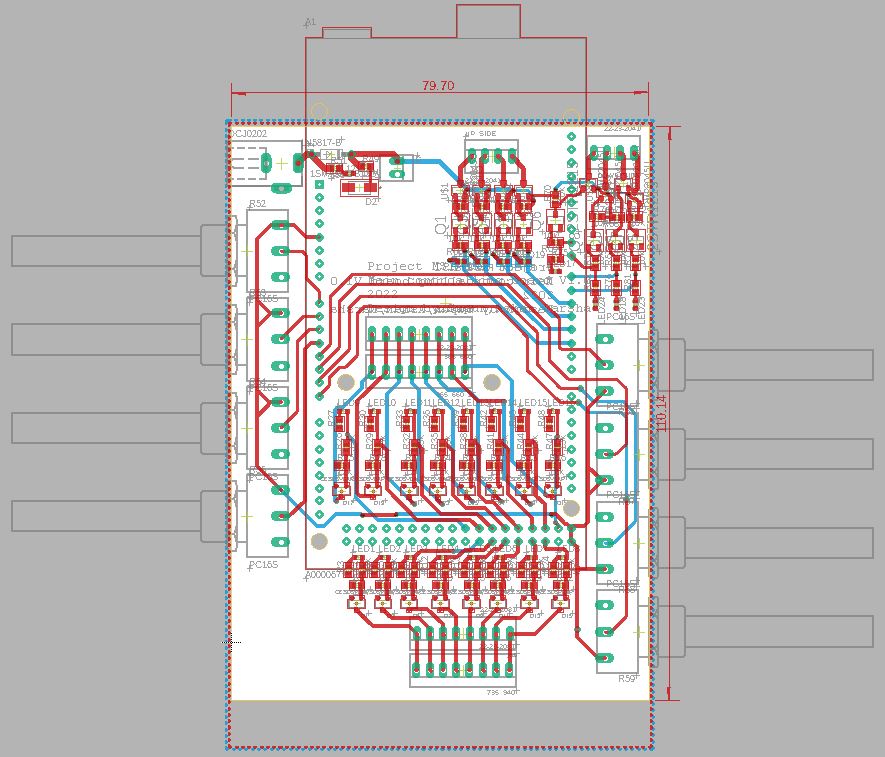}
\caption{PCB of the control board.}
\label{fig_pcb_2}
\end{figure}

Fig. \ref{fig_circuit}  shows the circuit of a single color band. The LEDs are connected to the power supply through two  MOSFETS.  One BJT is used to block the back current into the microcontroller.  In addition to these components, biasing resistors and current controlling resistors are used as necessary.  The microcontroller is used to provide the switching and PWM signals into the MOSFETS. The controller board is designed as a shield for the Arduino Due board, so it can be attached directly to the microcontroller board. The PCB layout of the control board is illustrated in Fig. \ref{fig_pcb_2}.

\subsubsection*{\bf Turning on/off the corresponding color band in the light panel.}

Switching on/off of a specific color band is handled by the Q2 MOSFET. It is an N-MOSFET. When a digital HIGH signal is fed into the gate terminal (VGS = 5V) of the Q2 MOSFET,  the MOSFET goes to the saturation region, closing the circuit. The MOSFET acts as an electronic switch.

\subsubsection*{\bf Controlling the intensity of each segment of the light panel.}

Intensity control of each segment is handled by the Q3 MOSFET. It is a P-MOSFET. The Q1 BJT is used to isolate the microcontroller from the rest of the circuit. This also acts as a signal amplifier. The PWM signal is fed into the base of the BJT. With the PWM signal at the base of the BJT, the collector voltage follows the PWM signal which is then fed into the gate of the MOSFET. Which results in a varying intensity of the LED according to the PWM value. Here both the  MOSFET and the BJT work in the linear region. The MOSFET was selected due to its ability to operate in higher frequency ranges than other power transistors.

\subsubsection*{\bf Communicating with the controller software.}

The system is handled via the software running on the computer, the user can control the device from the GUI. The user commands are sent through the controller software and executed using the microcontroller. Therefore, the communication between the microcontroller board and the software should be maintained.  

\subsection{Camera and the Image Sensor}
The camera installed in the system is a FLIR monochrome industrial machine vision camera (BFS-U3-13Y3M) with a PYTHON1300 image sensor and a `C-mount' lens mount.

The automatic corrections such as White Balancing, Gamma and Sharpening are disabled to have uniformity over images. Furthermore, parameters such as Gain and Exposure Time are kept constant during the imaging procedure.

\subsection{Controller Software and the Graphical User Interface}
\subsubsection*{\bf Control Architecture}

\begin{figure}[tb]
\centering
\includegraphics[width=\linewidth]{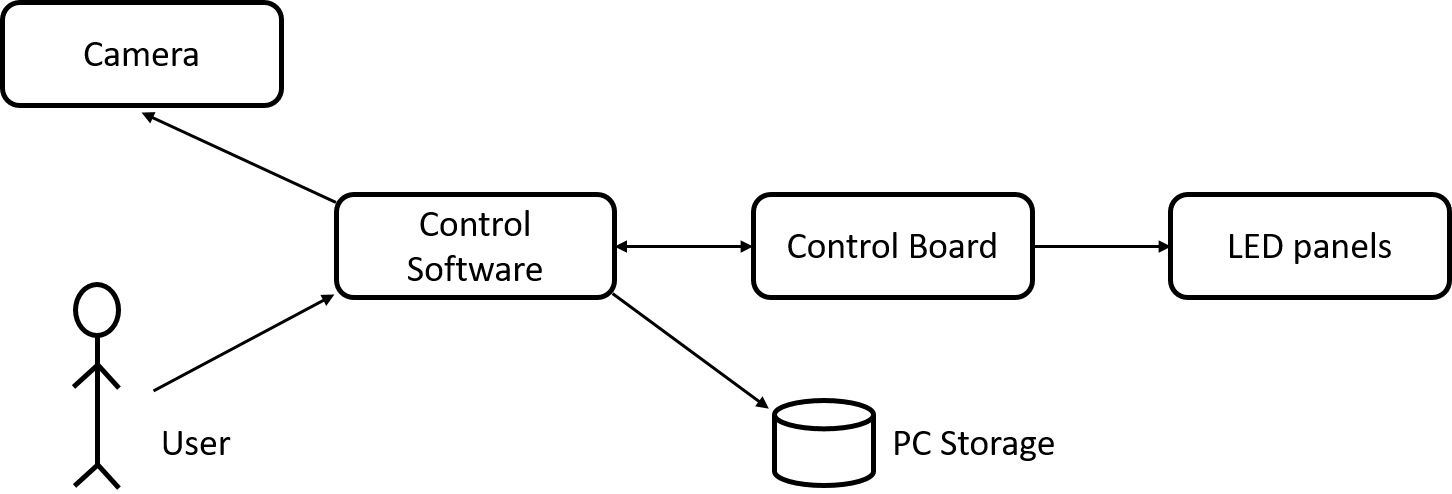}
\caption{Functional block diagram of the imaging system.}
\label{fig_block_diagram}
\end{figure}

The central control element of the system is the desktop application running on a computer.  Where it receives user inputs and then communicates to the camera and the control panel. The control panel turns the relevant LEDs on and immediately after, the camera captures the image. When the capturing ends, the LEDs are turned off instantly. A three-way handshake method is implemented to ensure the synchronization between the camera and the LEDs. The desktop app stores the captured images in the local disk of the computer in which the app is running. The communication structure is depicted in Fig. \ref{fig_block_diagram}.

\subsubsection*{\bf Control board firmware}
Arduino Due is utilized as the microcontroller board of the system. The microcontroller board communicates with the desktop app via Universal Serial Bus (USB) interfaces using Universal Asynchronous Receiver-Transmitter (UART). Microcontroller board firmware works in three states,
\begin{enumerate}
\item{Waiting.}
\item{turning on all the LEDs sequentially.}
\item{Turning on a given LED}
\end{enumerate}

Initially, the program is in the waiting state. Control characters are sent by the desktop app as a mechanism to switch the state. If a match is recognized, the state will switch from the waiting state to the corresponding state. At the end of each state, the program returns to the waiting state.

The light intensity of each light segment can be adjusted using 8 potentiometers on the board. Four potentiometers are dedicated to the top LED panel while the other four are dedicated to the bottom LED panel. The intensity of each LED panel varies with the input signals taken from the potentiometers. The program takes the reading of each potentiometer before proceeding to the state two or three. Then the PWM signals are generated by the program according to the potentiometer readings. Once the program is in state 1 or 2 (on/off stage of LEDs) the intensity cannot be changed.  

\subsubsection*{\bf Desktop application and camera control}

The desktop app controls the microcontroller board as well as the camera. Also, it interacts with the local disk of the computer. To access the camera, the Spinnaker SDK is used.


There are six indicators on the interface to show the intensity level of LEDs. As mentioned above, these levels change with the inputs of the eight potentiometers. The user can either capture all the color bands with one press of a button or they may select only the required color bands and capture them exclusively. While the image acquisition procedure is underway, each captured image is previewed on the GUI.

\subsection{Calibration and Image Acquisition}

\begin{figure}[tb]
\centering
\subfloat[\centering \textrm{Without corrections}]{{\includegraphics[width=\linewidth]{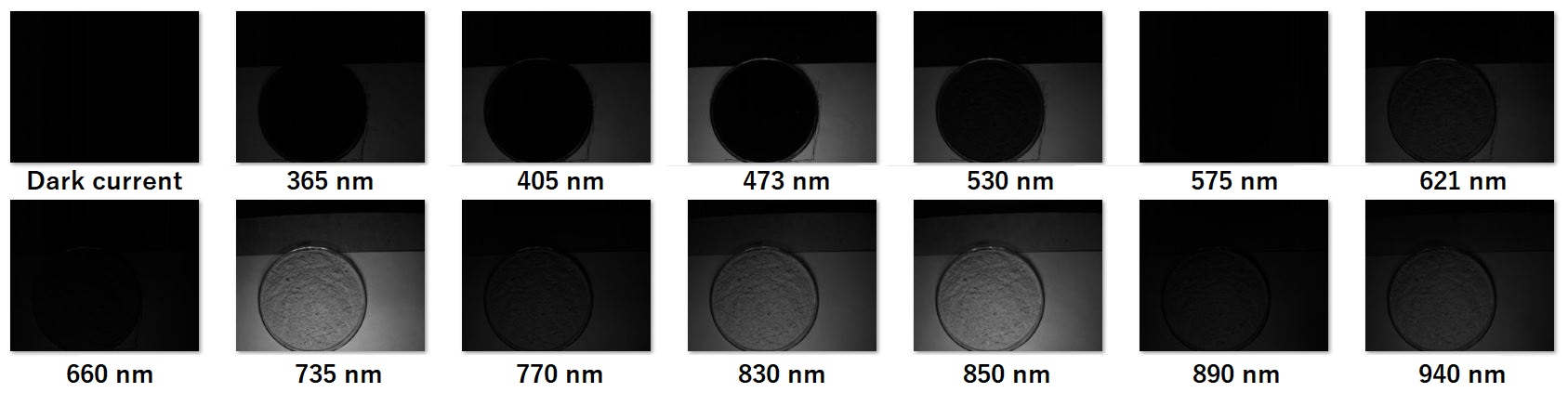} }
\label{fig_samples_uncorrected}}%
\qquad
\subfloat[\centering \textrm{With corrections}]{{\includegraphics[width=\linewidth]{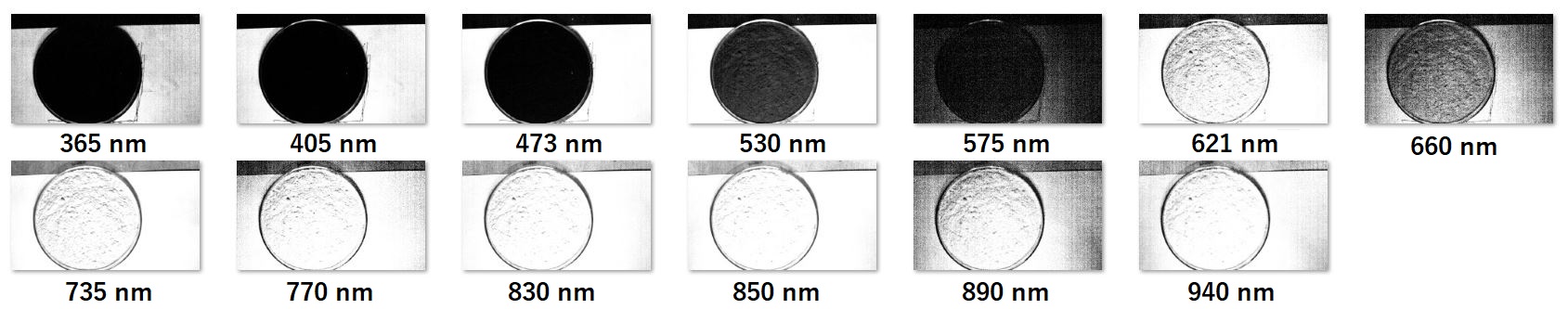} }
\label{fig_samples_corrected}}%
\caption{Captured images for each band with and without spatial-spectral corrections for case study 1.}
\label{fig_samples}
\end{figure}

To image a specimen using the proposed system, a certain procedure must be followed. This includes the mode selection and calibration. First, the prepared sample must be placed inside the image capture environment. After that, the required mode of imaging (reflectance or transmittance) must be selected using the control board. A key thing to consider here is that the platform which holds the sample must also be changed according to the operation mode. It must be opaque for the reflectance mode and it must be transparent for the transmittance mode. Then, the aperture of the camera must be adjusted. This setting controls the amount of light entering through the lens of the camera. For a sample that might saturate the sensor of the camera, a lower aperture is suitable. This will prevent the loss of information when imaging.  After that, the focus of the camera must be set. This ensures the best resolution and sharpness of details in the image. Next, the required color bands must be selected via the GUI. Finally, all the selected color bands of the specimen will be captured and stored in the computer as presented in Fig. \ref{fig_samples_uncorrected}.


\section{Materials and methods}

The reliability of the device as a measurement system was assessed by considering several metrics such as spatial consistency (reflectance intensity, spectral signature distortion), repeatability and variance of sample measurements. A single uniform piece of white paper was imaged using the device and the above metrics were observed.

In order to validate and evaluate the proposed imaging system, two methods were used. One was standard color testing and the other was validation by real experimental procedures and results.

For the standard color testing, a color chart with 24 distinct colors was used. Different algorithmic techniques were used to differentiate between colors using the spectral information captured using the system. 

As real experimental testing, two case studies were created to cover the full capability of the system.

\begin{enumerate}
\item{Estimation of adulteration level of turmeric powder with wheat flour \cite{Cite_18} (solid powdery specimen).}
\item{Estimation of adulteration level of coconut oil with palm oil (liquid specimen).}
\end{enumerate}

As per the case study 1, turmeric powder is a solid specimen which can be imaged in the reflectance mode while also being a type of sample which can then be dissolved in water and again imaged in transmittance mode. This case study provides the ground to validate both reflectance and transmittance modes while enabling it to create a dataset with one sample having both reflectance and transmittance data bundled together in the feature space increasing the dimensionality of the dataset. Thus, validates the proposed merged mode.  

As per the case study 2, coconut oil is a liquid specimen which can be imaged in the transmittance mode. From this case study, the performance of the transmittance mode of the system can be thoroughly analyzed and compared with other modes. The main objective is to build a functional mapping between the adulteration level and the transmittance spectral data. 

\subsection{Sample Preparation}

\subsubsection*{\bf Case Study 1: Turmeric Adulteration}

Samples of authentic turmeric powder, made from freshly harvested turmeric rhizomes, were gathered for the study. To adulterate the turmeric, the pure turmeric powder was then mixed homogeneously with varying amounts of rice flour, creating different ratios ranging from 0\% to 40\% by weight. For each adulteration level (0\%, 5\%, 10\%, 15\%, 20\%, 25\%, 30\%, 35\%, and 40\% by weight), nine identical samples were prepared, resulting in a total of 81 powdery samples.

In order to be used in the merged mode described above, new dissolved samples had to be created using the same powdery samples. For that, prepared turmeric powder samples were dissolved by mixing them with an equal volume of distilled water. The mixture was then stirred thoroughly until all the solid particles had completely dissolved in the water. A total of 81 dissolved turmeric samples were prepared as well.

\subsubsection*{\bf Case Study 2: Coconut Oil Adulteration}

This experiment is carried out to assess palm oil adulteration in coconut oil. The pure coconut oil was combined with varying proportions of palm oil, resulting in different volume ratios ranging from 0\% to 40\%. For each level of adulteration (0\%, 5\%, 10\%, 15\%, 20\%, 25\%, 30\%, 35\%, and 40\% by volume), eight identical samples were prepared, resulting in a total of 72 liquid samples.

\subsection{Imaging Using the System}

\subsubsection*{\bf Case Study 1: Turmeric Adulteration}

During the process of obtaining multispectral images from the adulterated turmeric powder samples, a meticulous procedure was followed. First, an equal amount of the powdered samples was spread evenly across the surfaces of the petri dishes. This meticulous distribution was essential to establish a uniform layer, ensuring consistency across all samples. Then, the reflectance mode of the device was utilized to capture images of the powder samples. First, placed turmeric powder-filled petri dish in the image-capturing environment, ensuring consistent placement throughout the experiment. In this process, the light was projected onto each sample, and the reflected light was captured using the camera. 

To employ the merged mode for turmeric analysis, it was necessary to capture the dissolved samples created from the powdery samples. For that,  the device was switched to transmittance mode. Transmittance images involve passing light through a substance and capturing the resulting information. The dissolved powder mixture was poured carefully into a container specifically designed with a transparent bottom and placed in the image-capturing environment where the transparent bottom of the container enabled the light rays to pass through the mixture unobstructed. Then, the light was transmitted through each sample and it was captured using the camera.

\subsubsection*{\bf Case Study 2: Coconut Oil Adulteration}

The coconut oil samples, after being prepared, were placed inside a container with a transparent bottom. The mixture was then gently stirred to ensure that no air bubbles remained in the mixture. This careful stirring helped create a consistent and even composition. Using the transmittance mode of the device, the light was passed through each sample and it was captured as an image. Captured sets of images were used for further analysis.

\subsection{Multispectral Image Preprocessing}

Image preprocessing is a vital step prior to multispectral image analysis because it extracts the useful information and enhances it. Preprocessing can also reduce the effect of noise which may have occurred during image acquisition.

\subsubsection*{\bf Image Cropping}

Cropping is a common image preprocessing technique that removes unwanted elements to direct the focus onto specific areas of interest. It is often performed before further processing to eliminate distractions and to ensure only necessary information is used for further analysis. Therefore, every captured image was cropped into a 100×100-pixel image which only contains the information about the specimen to be analyzed. Since there are 14 images of a single sample (consisting of a dark image and 13 spectral bands), an identical region was cropped from all 14 images.

\subsubsection*{\bf Dark Current Reduction}

MSI systems face random noise sources like camera read-out, wire connections, data transfer, electronic noise, and analog-to-digital conversion, which can affect the outcomes of image analysis \cite{els_4}. In the preprocessing stage random noise is reduced using dark current subtraction. Pixel recording can happen even when any source of light is not present. Dark current primarily emerges from currents generated during the creation of the depletion region and irregularities on the silicon lattice surface of the photodiode. To mitigate this effect, a dark current image is captured at the beginning of multispectral imaging, and then it is subtracted from all the following lighted images. This technique reduces random noise in photodiodes and photosensors.

\begin{figure}[tb]
\centering
\subfloat[\centering Spectral signatures without normalization]{{\includegraphics[width=0.7\linewidth]{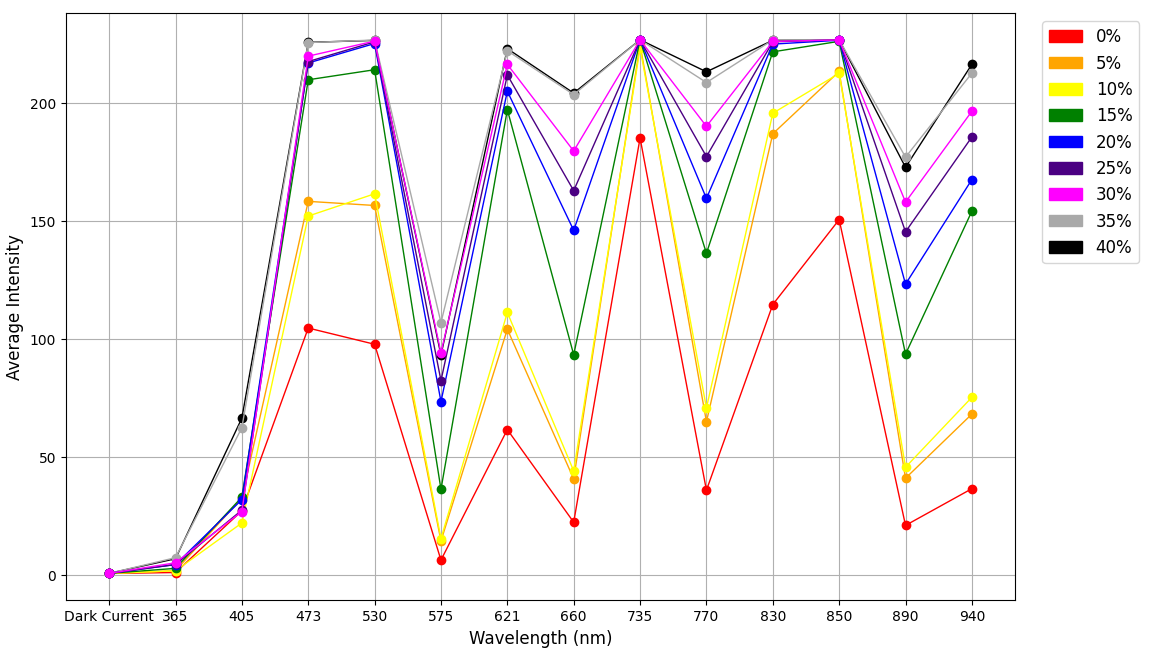} }
\label{fig_spec_sig_raw}}%
\qquad
\subfloat[\centering Normalized spectral signature]{{\includegraphics[width=0.7\linewidth]{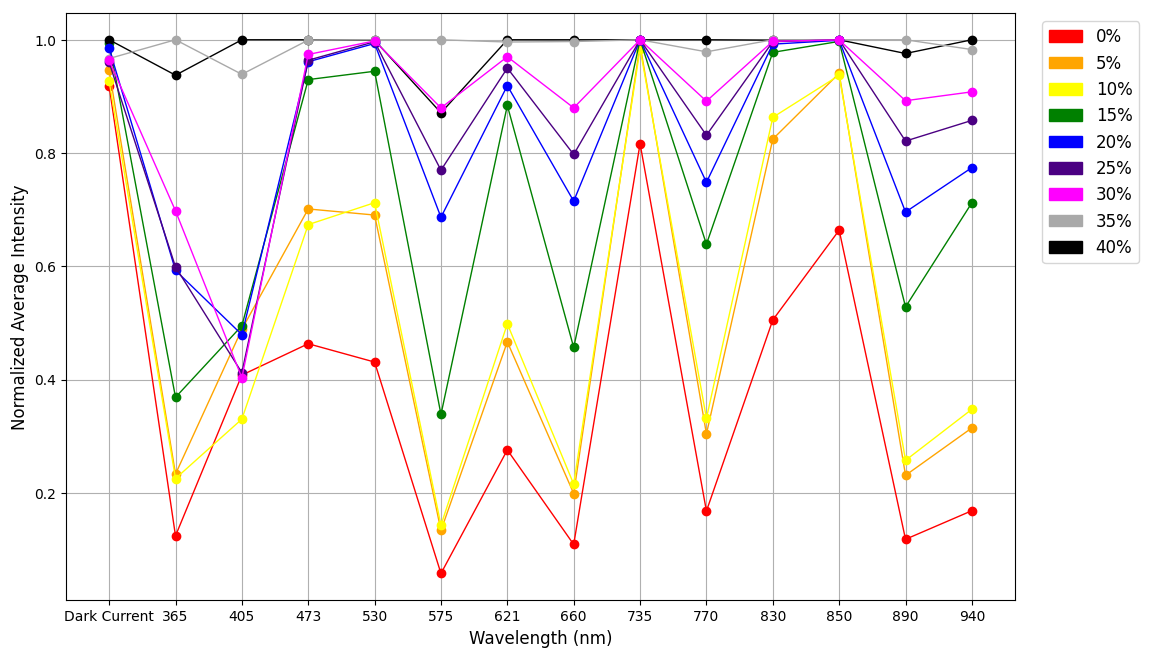} }
\label{fig_spec_sig_corrected}}%
\caption{Spectral signatures of coconut oil samples for case study 2 with and without spectral band normalization.}
\label{fig_oil_spec_sig}
\end{figure}


\subsubsection*{\bf Image Correction}

Due to nonuniformities of illumination due to light source placement, image sensor detection differences across wavelengths and different illumination intensities of LEDs, corrections to the captured images are necessary before being used in the analysis to remove any device-generated bias. Several steps in the analysis procedure are dedicated for this nonuniformity mitigation. Data normalization, feature space generation by dimension reduction and training procedure compensate for most of the data nonuniformities in images. In addition to that, post-imaging spatial and spectral corrections are done to improve data consistency.

Spatial correction is done to flatten the intensity response of the image over the imaging plane. This is achieved by assessing the reflectance intensity map of each color band and constructing suitable spatial correction functions from them.

Spectral correction is done to obtain a spectrally balanced set of multispectral images regardless of the characteristics of the image sensor and illumination LEDs. For that, an experimental procedure is used to obtain the spectral characteristic function of the device across color bands. Then corrections are performed for the required color bands.

As shown in Fig. \ref{surface_plot}, the slope-shaped bias present in the raw image is corrected by applying spatial corrections. Then, the spatial corrected image is subjected to spectral correction which is also shown in Fig. \ref{surface_plot}. A set of captured images with and without corrections is shown in Fig. \ref{fig_samples}.

\subsubsection*{\bf Bilateral Filtering}

Bilateral filtering is another noise reduction preprocessing technique. It is utilized to reduce noise while preserving edges and fine details. The bilateral filter calculates the weighted average of neighboring pixels, where the weights are determined by both spatial proximity and intensity similarity. This weighting scheme helps to smoothen the image while preserving edges.

\subsection{Representation of Multispectral Image Data}

\subsubsection*{\bf The Data Matrix} \label{subsection_data_mat}

Multispectral data consists of monochromatic images representing intensities at different wavelength bands. These images contain vector pixels with spectral and spatial information. Initially, from all the images,  $10\times10$-pixel sections were averaged out to create superpixels which are less noisy.  This was done for cropped images with the size of $100\times100$-pixels, resulting in $100$ superpixels per image. Said superpixel intensities were stored vertically in a data matrix - $X$ with dimensions of $100\times13$ (13 spectral bands). This process was repeated for all the samples, resulting in a vertical stack of intensity values of superpixels.

For the merged mode, the reflectance data matrix  { $R$} for a sample is generated by vertically stacking image pixels through the 13 columns corresponding to the 13 color bands of the device. For a single color band image having the dimensions of $m \times n$, the total number of pixels in that image is $\alpha = m\cdot n$. Therefore, the resulting reflectance data matrix {$R$} is an $\alpha \times 13$ matrix. Similarly, for the same sample, the transmittance data matrix {$T$} is also $\alpha \times 13$. Then, the matrix {$R$} and matrix {$T$} are joined horizontally producing the merged data matrix {$M$} for the said sample. {$M$} has the dimensions of $\alpha \times 26$. If there is a total of {$p$} number of samples, the final data matrix will be an $(\alpha\cdot p) \times 26$ matrix.

\subsubsection*{\bf Spectral Signatures}

After creating the data matrix, the average pixel intensity of all the samples in each adulteration level across every spectral band was obtained. Then, the average pixel intensity vs wavelength was plotted for different adulteration levels. The spectral signature plots of Coconut oil samples are illustrated in Fig. \ref{fig_oil_spec_sig}. The raw spectral signatures without any modifications are shown in Fig. \ref{fig_spec_sig_raw} while in Fig. \ref{fig_spec_sig_corrected}, spectral band normalization is performed to minimize the induced sensor and LED nonuniformities.

\begin{figure}[tb]
\centering
\includegraphics[width=0.7\linewidth]{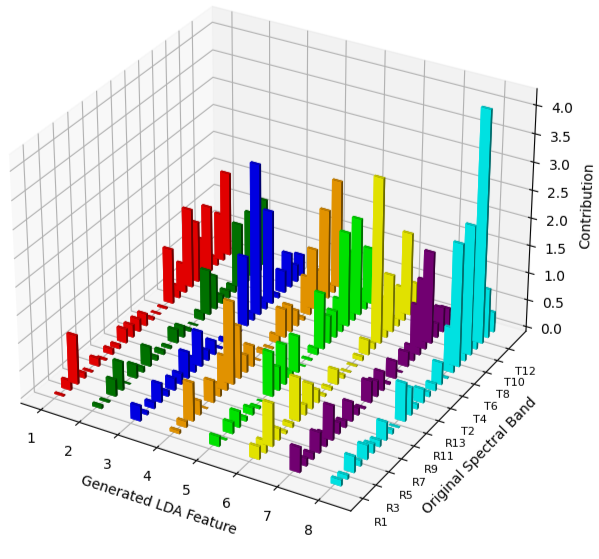}
\caption{Contribution of each transmittance and reflectance spectral band for optimal feature generation in case study 1.}
\label{fig_eigen_plot}
\end{figure}


\subsection{Feature Extraction and Analysis}

PCA and LDA are the main dimensionality reduction and feature extraction methods used in this study. PCA is used as a dimensionality reduction technique where it transforms the feature space into a new uncorrelated feature space containing all the useful information \cite{review_2_pca}. LDA is another dimensionality reduction technique that focuses on maximizing the separability of classes as a supervised method \cite{review_2_lda}.

Construction of the dataset as mentioned in Section \ref{subsection_data_mat}, allows the dimension reduction techniques such as PCA and LDA to generate new case-specific uncorrelated optimal features as linear combinations of original spectral bands which then is used in classification by ML and AI modules. Fig. \ref{fig_eigen_plot} depicts the contribution of each reflectance and transmittance spectral band to the optimal set of uncorrelated and orthogonal features via LDA for case study 1.

\subsubsection*{\bf Standard Color Testing}


The images acquired for the standard color palette consisting of 24 distinct colors were split into non-overlapping 75\%-training and 25\%-testing sets. PCA and LDA were separately applied on the training set to reduce the dimensionality of the dataset and extract features for the ML classifiers. Then, the classifiers were validated using the testing set which is projected to the reduced dimensions.


\subsubsection*{\bf Case Study 1: Turmeric Adulteration}

The dataset was split into non-overlapping 75\%-training and 25\%-testing sets. LDA was applied to the training set before classification models were trained to classify samples based on their adulteration levels in three stages and they were validated using the test set. As mentioned in the sample preparation, relevant adulteration percentages were selected as the class labels. First, samples were classified using only reflectance mode images. Then, they were classified using only transmittance mode images. Finally, classification was done by combining both reflectance and transmittance images (merged mode).

\subsubsection*{\bf Case Study 2: Coconut Oil Adulteration}

The dataset was split into non-overlapping 75\%-training and 25\%-testing sets. LDA was applied to the training set before the analysis was carried out in two methods. The first method was the coconut oil sample classification using different ML classifiers based on their adulteration levels where the adulteration levels are taken as the class labels. The classifiers were validated using the test set.

In the second method, KL divergence was used to develop a functional map between adulteration levels and concentrated MSI data. A statistical measure from information theory known as the KL divergence metric is frequently used to quantify the deviation of one probability distribution from a reference probability distribution. The following equation is used to calculate KL divergence.

\begin{equation}
K L(P \| Q)=\sum_{x_i} P\left(x_i\right) \log \frac{P\left(x_i\right)}{Q\left(x_i\right)}  
\end{equation}

Q(x) is the sample probability distribution, while P(x) is the reference probability distribution. The KL divergence was calculated for nine replicates for each adulteration level using their MSI data. The 0\% adulteration level was used as the reference.

\section{Results}

\subsection{Reliability Assessment of the Measurement System}

\begin{figure}[tb]
\centering
\includegraphics[width=\linewidth]{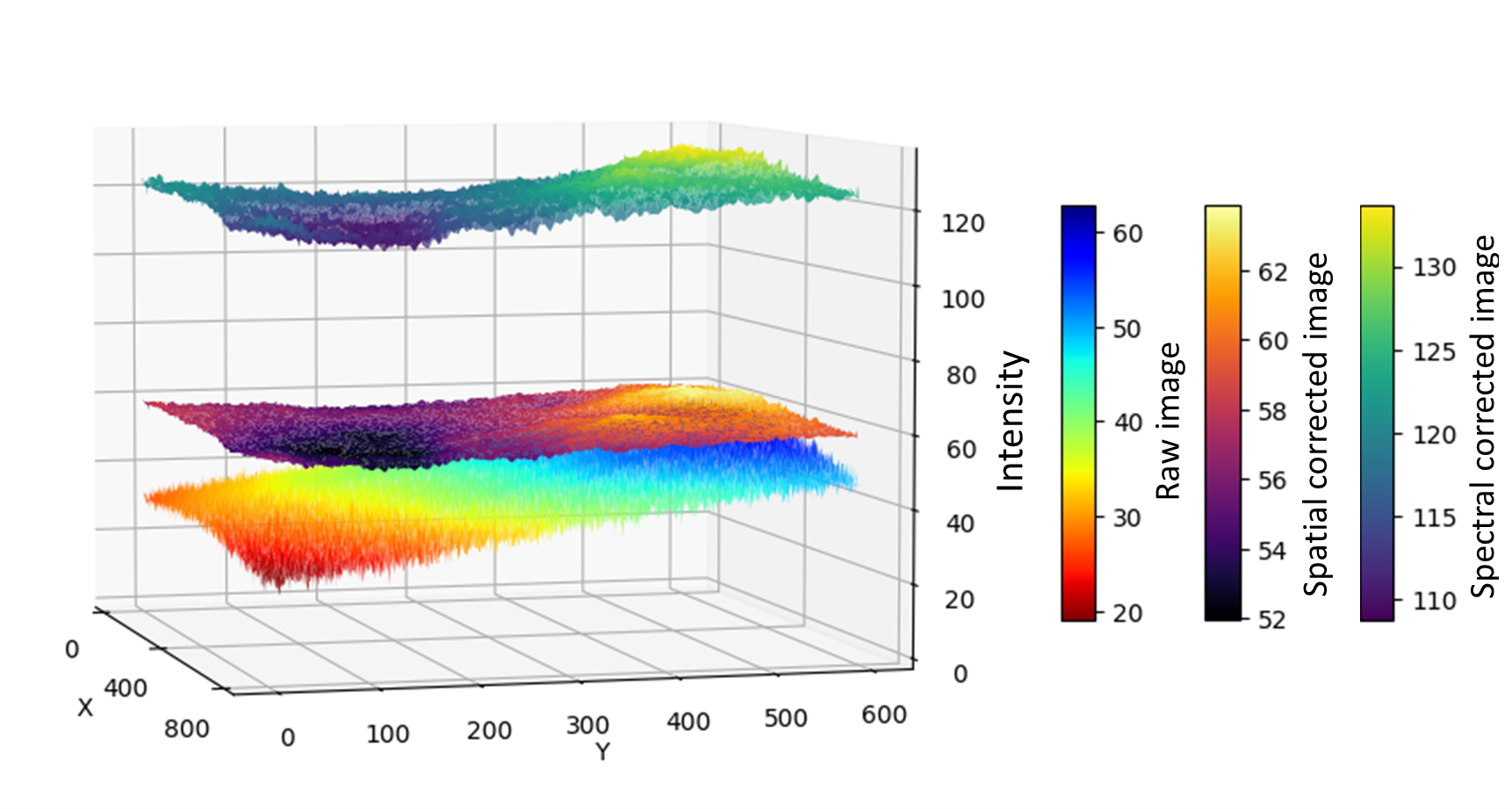}
\caption{Spatial distribution of normalized reflectance intensity of a white paper illuminated by 770 nm band. Without corrections, after spatial corrections, after spatial-spectral corrections}
\label{surface_plot}
\end{figure}

\begin{figure}[tb]
\centering
\subfloat[\centering \textrm{Without corrections}]{{\includegraphics[width=0.4\linewidth]{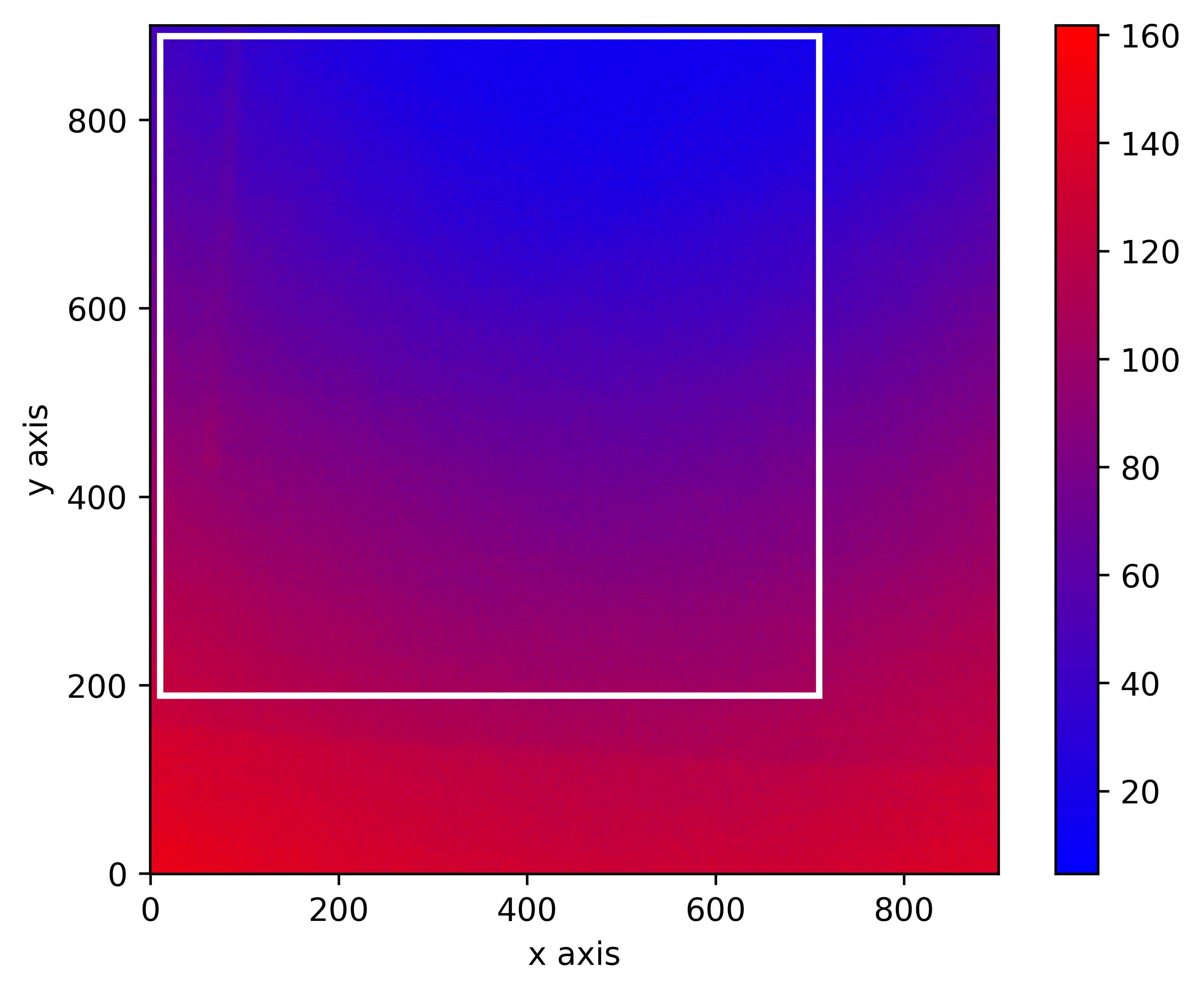} }
\label{heatmap_before}}%
\qquad
\subfloat[\centering \textrm{With corrections}]{{\includegraphics[width=0.4\linewidth]{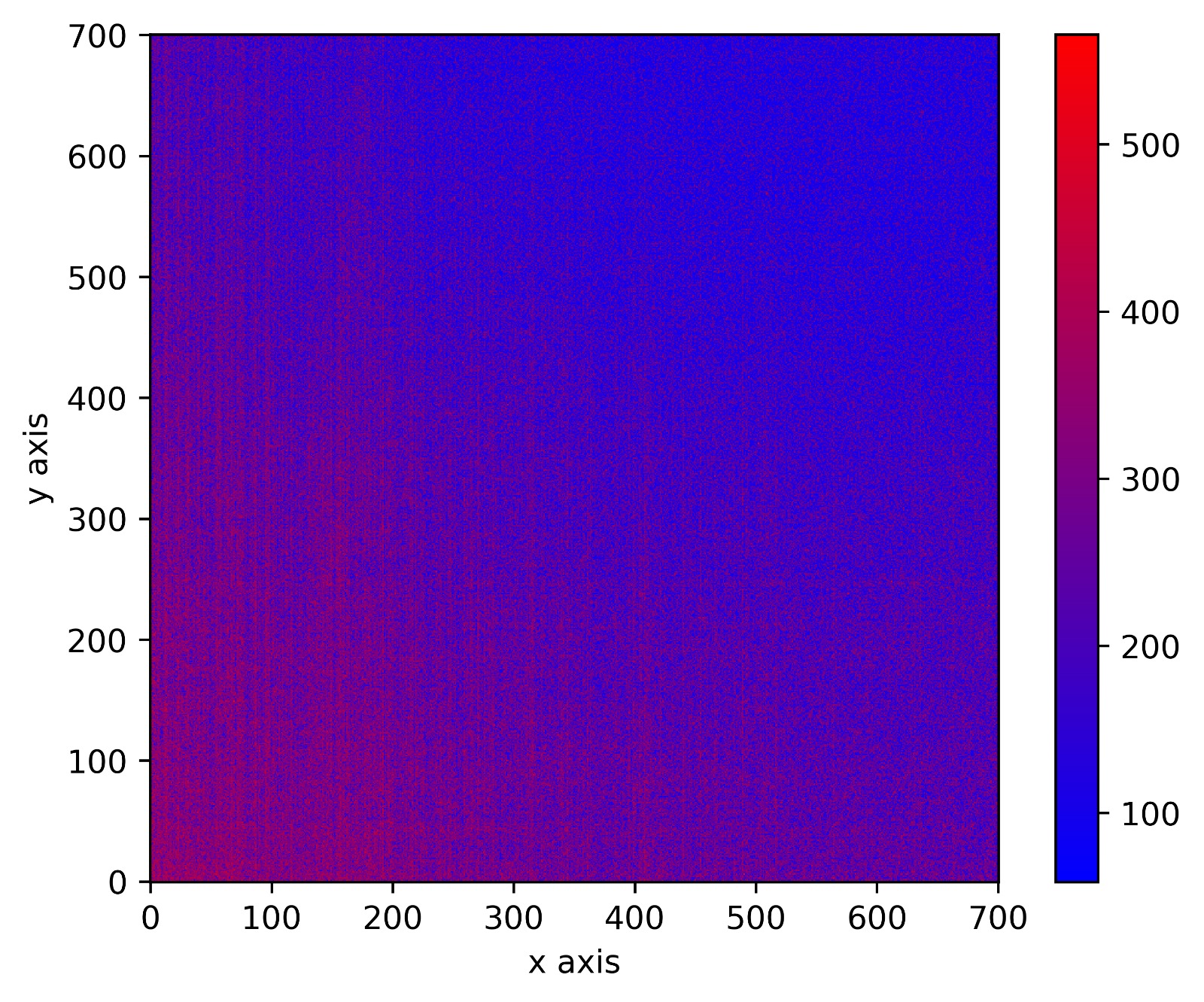} }
\label{heatmap_after}}%
\caption{Spatial variation of the Euclidean distance of the spectral signatures with respect to the center pixel of the highest illumination region when imaging a white paper with and without spatial-spectral image corrections. The highlighted region is extracted for corrections.}
\label{heatmap_plot}
\end{figure}

\begin{figure}[tb]
\centering
\includegraphics[width=0.7\linewidth]{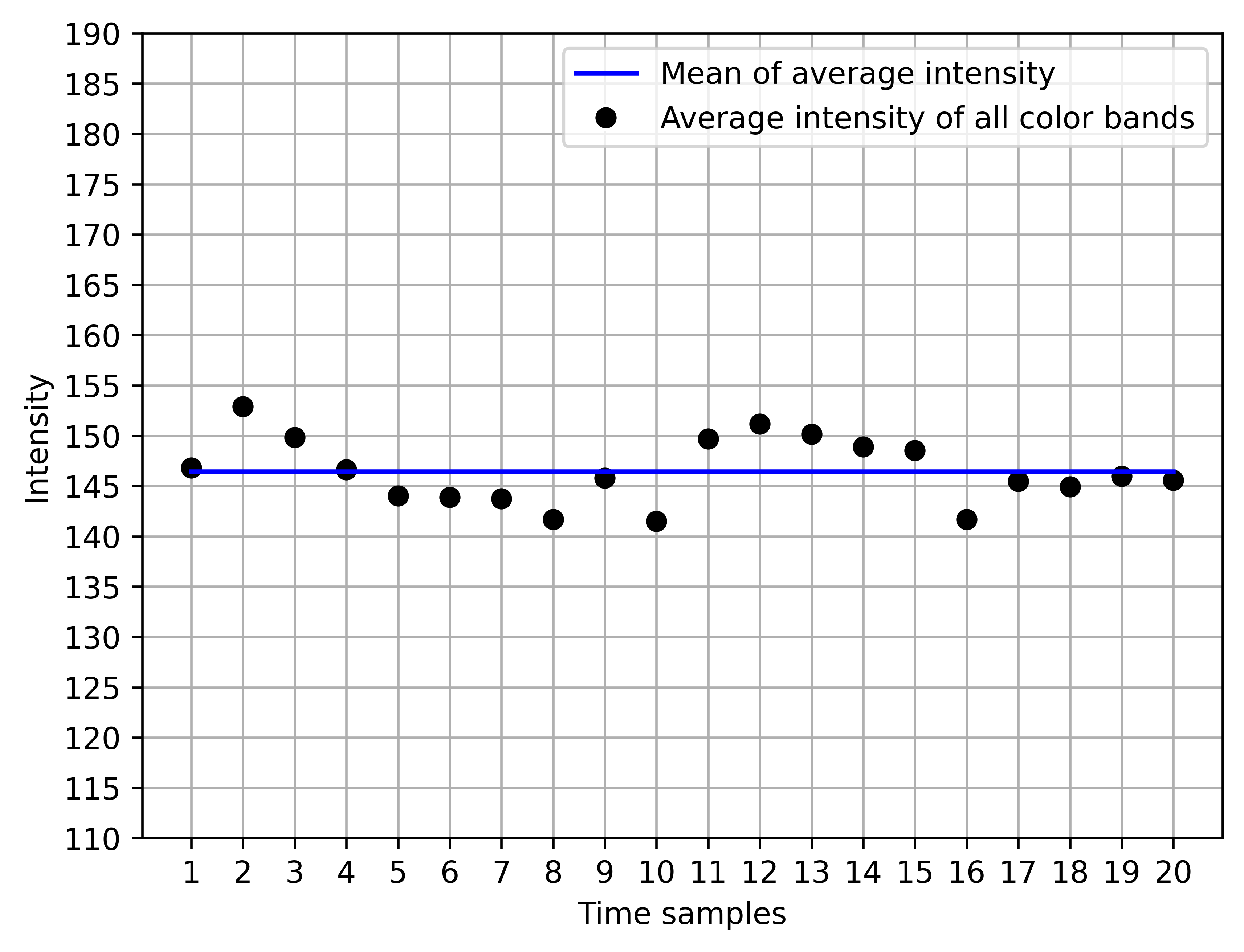}
\caption{Temporal variation of the overall intensity of all color bands.}
\label{repeat_plot}
\end{figure}

Reflectance spatial consistency of the captured images was ensured by the spatial corrections as shown in Fig. \ref{surface_plot}. Same shape of reflectance intensity was observed for all the color bands. However, only the intensity plot of  770 nm wavelength is displayed for brevity. In addition to that, the spectral signature distortion with the spatial difference is depicted in Fig \ref{heatmap_plot}. Post-imaging spatial-spectral corrections have significantly reduced the spectral signature distortion as observed by Fig. \ref{heatmap_after} compared to Fig. \ref{heatmap_before}.  The best location for specimen placement in the system would be the intersection between the highest intensity area of Fig. \ref{surface_plot} and the lowest spectral signature distortion area of Fig. \ref{heatmap_plot}. The repeatability of the system was tested by imaging the same sample with temporal gaps. As indicated in Fig. \ref{repeat_plot}, MSI parameters across all the color bands did not show a substantial deviation with time. The maximum percentage deviation from the mean with time was observed to be 4.41\%. In the functional mapping development attempt in case study 2, box plots were utilized to analyze the distribution of data points of the same class created using KL divergence as displayed in Fig. \ref{fig_oil_regress}. The data points of the same class were observed to be placed without a high variance indicating the validity of the measurements.

\subsection{Standard Color Testing}

\begin{figure}[tb]
\centering
\includegraphics[width=0.7\linewidth]{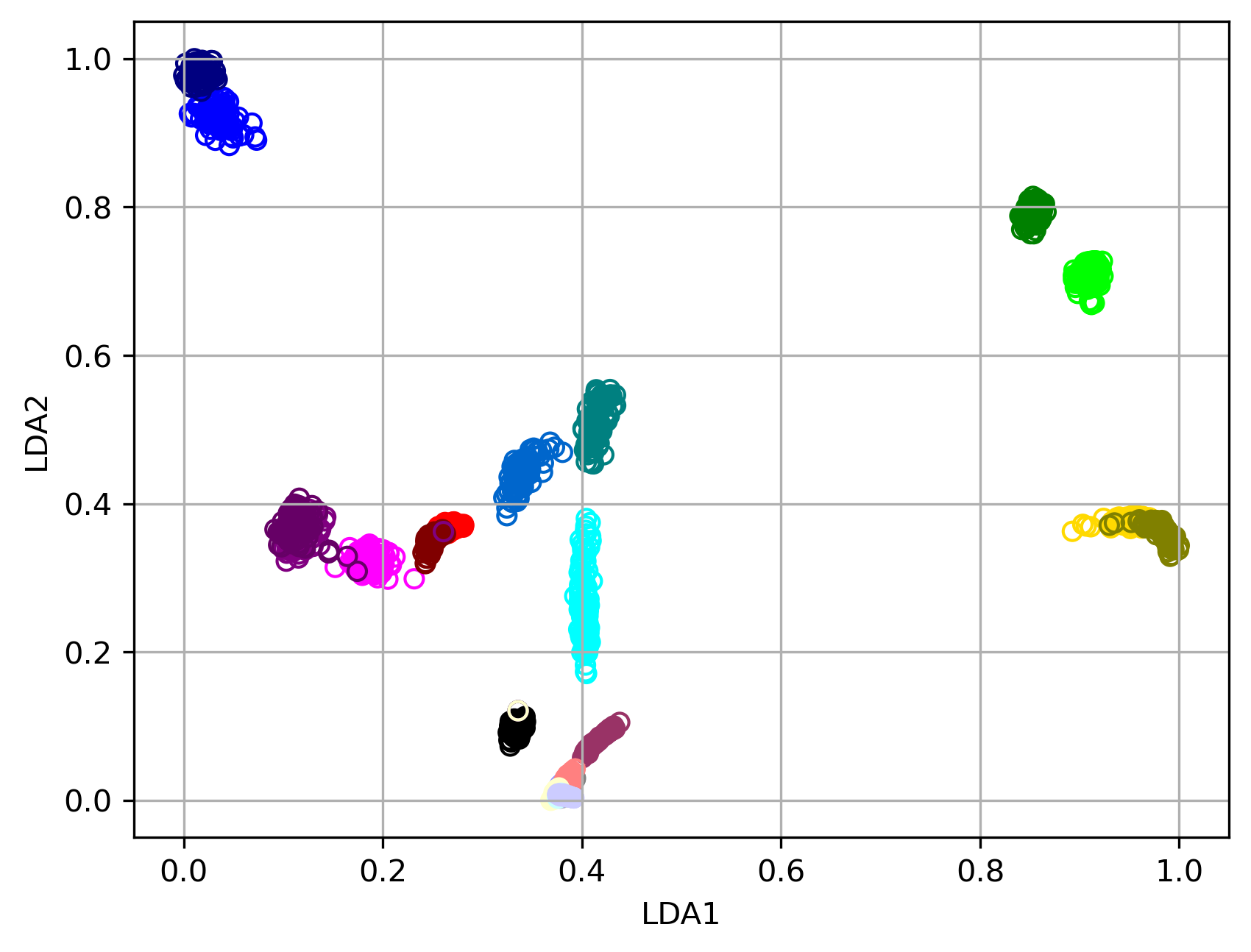}
\caption{ Standard color palette class distribution on normalized LDA dimensions.}
\label{fig_color_lda}
\end{figure}

 After preprocessing and dimension reduction were done for the standard color palette, Fig. \ref{fig_color_lda} shows the variation of data along the first two normalized LDA components. For the classification task, the Support Vector Machines (SVM) Classifier coupled with PCA and LDA provided the best classification accuracy of around 90\% when identifying individual colors of the standard color palette using their MSI parameters. Classification accuracies for the standard color test are listed in the TABLE \ref{tab_color_accuracy}.

\begin{table}[tb]
\centering
\caption{Standard color palette classification accuracy.}
\begin{tabular}{lll}
Classifier            & With PCA  & With LDA  \\
\hline
Decision   tree       & 0.86      & 0.88  \\
KNN                   & 0.89      & 0.88  \\
Logistic   regression & 0.84      & 0.84  \\
Random   forest       & 0.89      & 0.91  \\
SVM                   & 0.93      & 0.90
\end{tabular}
\label{tab_color_accuracy}
\end{table}

\subsection{Case Study 1: Turmeric Adulteration}

\begin{figure}[tb]
\centering
\includegraphics[width=\linewidth]{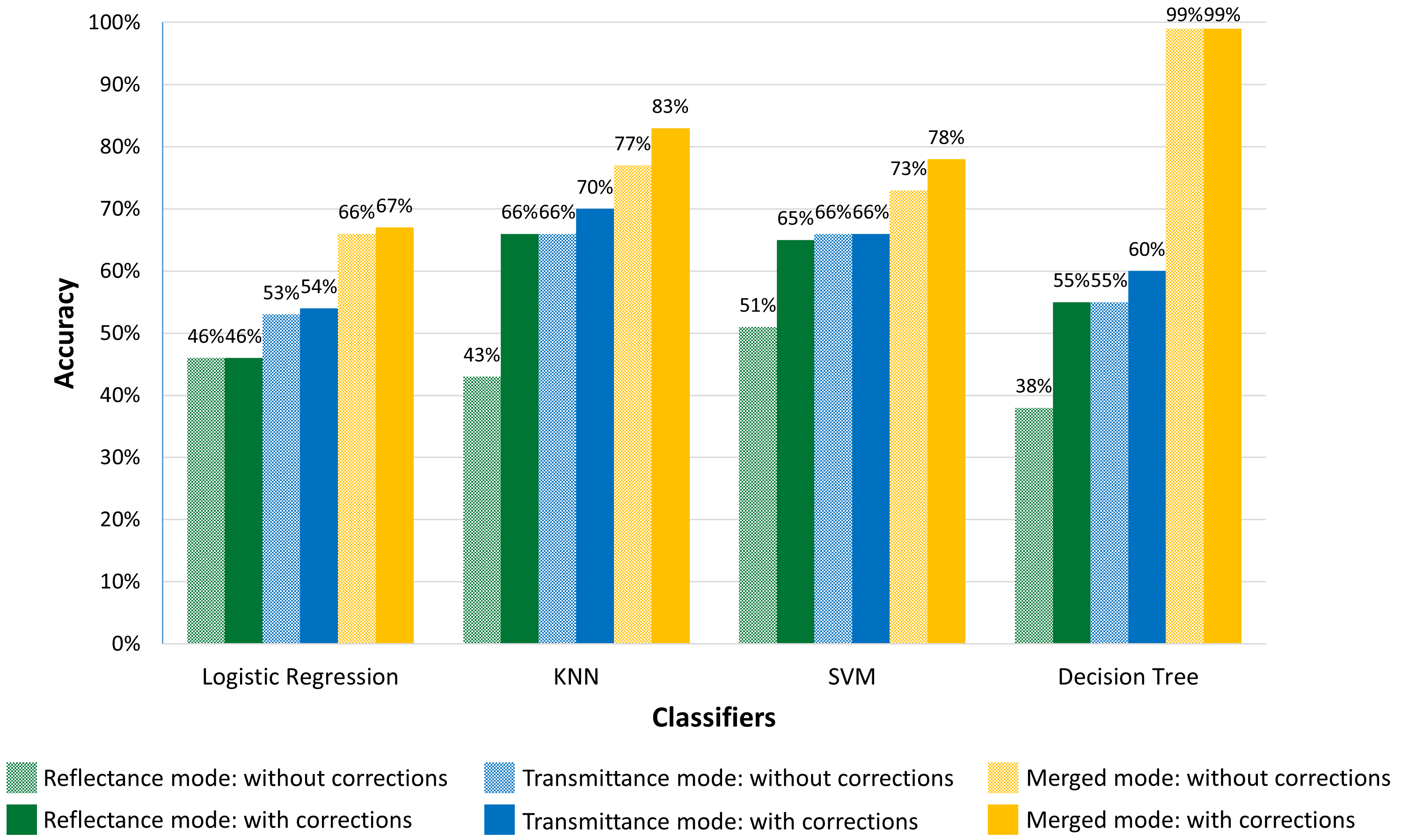}
\caption{Case Study 1: Turmeric classification accuracy across operational modes and ML classifiers with and without spatial-spectral corrections.}
\label{fig_turmeric_accuracy}
\end{figure}

On the case study pertaining to turmeric adulteration level classification, new LDA features which are specifically generated for this case study as linear combinations of original merged mode spectral bands are depicted in Fig. \ref{fig_eigen_plot}. It is observed that transmittance spectral bands have contributed more toward new LDA features compared to reflectance spectral bands for this specific case study. Out of the three classification stages, the best accuracy recorded for the reflectance-only classification was 66\%  given by the K Nearest Neighbors (KNN) classifier. For the transmittance-only classification, the accuracy slightly increased compared to the reflectance-only stage with 70\% being the highest. The classifiers in the combined reflectance-transmittance (merged mode) stage yielded the best accuracies for all the classifiers with the Decision Tree Classifier providing the best accuracy of around 99\%, thereby establishing the validity of the newly proposed merged mode as an effective method to analyze different specimens. Furthermore, the results have increased in accuracy when post-imaging spatial-spectral corrections were performed in almost all the cases. Fig. \ref{fig_turmeric_accuracy} depicts the accuracy comparison between modes and classifiers with and without corrections.

\subsection{Case Study 2: Coconut Oil Adulteration}
\subsubsection*{\bf Classification model}

Among the employed classifiers to classify coconut oil samples based on their adulteration level, the Decision Tree classifier provided the best accuracy around 95\%. All the algorithms yielded very good accuracies which are presented in TABLE \ref{tab_oil_accuracy}. All of them were greater than 88\%.  In conclusion to this case study, it can be verified that the proposed MSI system provides a good solution to quantify the adulteration level of palm oil in coconut oil. This was made possible by the dedicated transmittance mode of the device and this establishes the viability of transmittance multispectral imaging to analyze liquid samples.

For transparent liquids such as coconut oil, some spectral bands tend to be near saturated as shown in Fig. \ref{fig_spec_sig_raw}. Therefore, additional spectral corrections other than spectral band normalization may be detrimental resulting in a classifier performance degradation. However, as can be noted from the results shown in TABLE \ref{tab_oil_accuracy}, the few remaining nonsaturated bands have provided sufficient information for classification as evident by the high levels of classification accuracy. Hence, the standard spectral band normalization depicted in Fig. \ref{fig_spec_sig_corrected} is seen to be sufficient for this particular problem.

\begin{table}[tb]
\centering
\caption{Case study 2: Coconut oil adulteration classification accuracy.}
\begin{tabular}{lll}
Classifier            & Accuracy \\
\hline

Logistic Regression & 0.89 \\
KNN                   & 0.95 \\
SVM                   & 0.95 \\
Decision Tree       & 0.95

\end{tabular}
\label{tab_oil_accuracy}
\end{table}

\subsubsection*{\bf Functional mapping}

\begin{figure}[tb]
\centering
\includegraphics[width=0.7\linewidth]{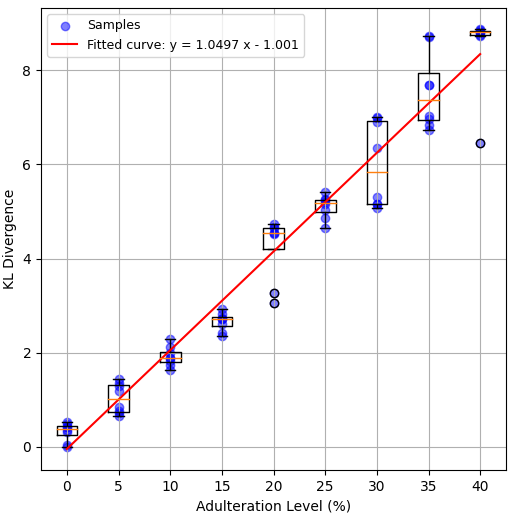}
\caption{Functional Mapping of Coconut oil adulteration level with KL Divergence from the pure sample.}
\label{fig_oil_regress}
\end{figure}

The constructed functional relationship between the KL divergence metric and the adulteration level of coconut oil is shown in Fig. \ref{fig_oil_regress}. The mathematical expression of the functional map is as follows,

\begin{equation}
{Y=1.0497 X-1.001}
\end{equation}

with R\textsuperscript{2} = 0.9558. Where X and Y represent the percentage adulteration level and the KL divergence respectively. As represented in Fig. \ref{fig_oil_regress}, the KL divergence increases with the level of adulteration. As a result, using multispectral imaging, the developed model can predict the palm oil adulteration level in coconut oil with significant precision.

\section{Conclusion}

A versatile multispectral imaging device that can be used for reliable assessment of food quality in a field setting is proposed in this paper. This system provides the capability of both reflectance mode MSI and transmittance mode MSI, with the ability to easily switch between the two modes. This enables the user to make the measurements using only transmittance, or only reflectance, or using both modes for the same sample preparation, or using both modes with different preparations derived from the same sample. In the case when both reflectance and transmittance images are acquired, it is possible to achieve better results by merging the measurements to obtain a higher number of features. For the instrumentation setup, an extensive spatio-temporal variation study was performed to identify the optimal area for imaging acquisition for processing. Different ML algorithms are used to classify food samples while KL divergence based signal processing algorithms were developed to obtain a functional relationship between measured samples and adulteration levels. The measurement attributes such as repeatability, consistency, precision, and accuracy of this device are assessed in detail via controlled calibration tests and case studies. Design features of this system include symmetric and modular light panels, a centralized controller module, reflecting hemispheres, and an intuitive GUI. 

The system as a whole was validated thoroughly by standard color palette testing as well as real experimental testing including food specimens such as turmeric powder and coconut oil. The system was able to identify different colors with all the classifiers providing more than 85\% accuracy while the SVM classifier was the best with 90\%. The transmittance mode was proved to be accurate with classifiers applied on coconut oil samples giving more than 89\% accuracy where KNN, SVM and Decision Tree algorithms yielded an accuracy of 95\%. The effectiveness of the merged mode was affirmed by the fact that the Decision Tree Classifier was able to provide the best classification results under the merged mode analysis utilizing powdered and dissolved turmeric samples. Reflectance mode and transmittance mode individually provided 38\% and 55\% accuracies respectively while merged mode was able to provide an accuracy of 99\%. The linear functional map developed in the regression model for determining the adulteration level of coconut oil can be considered successful as the obtained R\textsuperscript{2} value is 0.9558.  

While the paper is focused on food quality estimation, the applications of the device are not limited to food and can be extended further. The developed system can be used as a portable imaging device for any solid or liquid specimens. However, it is limited by the physical parameters of the specimen due to the dimensions of the imaging environment. It can be upgraded to contain specimens of different sizes and weights while retaining the mobility of the device. Although the system covers a reasonable range of the electromagnetic spectrum, compared to more expensive equipment, the imaging spectral range can be widened and the spectral resolution can be increased by adding more color bands. Following that improvement, the image sensor may be upgraded to have a flatter and wider sensitivity across the spectrum. Due to the lack of transmittance mode imaging studies, further research is required to develop transmittance spectral correctional functions where they would not contribute to the loss of information by saturating information carrying spectral bands of transparent liquids. The proposed merged mode strategy for higher dimensionality in feature space construction and KL divergence type metric utilization for obtaining functional relationships with adulteration and contamination levels can be extended beyond the recommended food quality applications.

\section*{Acknowledgments}
The authors would like to thank H.M.P.S. Madushan, K.K. Abewickrama, J.A.S.T. Jayakody and G.A.S.L. Ranasinghe of University of Peradeniya, Sri Lanka for their valuable contribution towards experimental testing of the system.

\bibliographystyle{IEEEtran}
\bibliography{Citations}

\vfill

\end{document}